\documentstyle[amssymb,epsfig]{aipproc}

\begin{document}


\title{Optimizing the Superlens:
manipulating geometry to enhance the resolution}

\author{Viktor A. Podolskiy$^{\dagger}$\thanks{E-mail: vpodolsk@physics.orst.edu;
WWW: http://www.physics.orst.edu/$\sim$vpodolsk}, Nicholas A. Kuhta$^{\dagger}$, and Graeme W. Milton$^{\ddagger}$}

\address{$^{\dagger}$ Physics Department, 301 Weniger Hall, Oregon State University\\
Corvallis OR 97331\\
$^{\ddagger}$Department of Mathematics, University of Utah\\
Salt Lake City UT 84112}

\maketitle

\begin{abstract}
We analyze the performance of a planar lens based on realistic negative index material in a generalized geometry. We demonstrate that the conventional superlens design (where the lens is centered between the object and the image) is not optimal from the resolution point-of-view, develop an analytical expression for the resolution limit of a generalized lens, use it to find the optimum lens configuration, and calculate the maximum absorption practical nearfield superlenses may have. We demonstrate that in contrast to the conventional superlens picture, planar imaging is typically accompanied by excitation of surface waves at both interfaces of the lens. 
\end{abstract}


Research on the properties of negative refractive index materials (NIMs)\cite{veselago} is among the most rapidly developing topics in modern science that may potentially lead to a number of unique applications including high-performance imaging and lithographic systems, new types of radars, and transmission lines\cite{pendry,sridhar,trLine,podolskiyPlanar,zhang,park}. One of the most promising applications of NIMs involves the use of a planar slab of negative refraction material as an optically-perfect imaging instrument\cite{pendry}, known as a superlens. The physics behind the operation of a superlens and the laws governing the resolution limits of this unique system have instigated considerable controversy\cite{pendry,vesperinas,pendryComment,vesperinasAnswer}. The recent analytical\cite{smithRESOLUT,merlin,webbSuperlens,stockmanSuperlens,podolskiyResolut}, numerical\cite{shvetsSuperlens,soukoulis}, and experimental\cite{zhang,shvetsSuperlens,soukoulis} results demonstrate that while a NIM-based system may outperform conventional (phase) lenses, its resolution {\it logarithmically} depends on material absorption, limiting all practical applications of superlens to the near-field zone\cite{podolskiyResolut}. However, existing analytical results describing the superlens performance, are limited to a single lens geometry, when a slab of negative-n (or negative-$\epsilon$) material is centered between an object and its image\cite{pendry} and the effects of the superlens design on its resolution are yet to be understood. 

\begin{figure}[t] 
\centerline{\epsfig{file=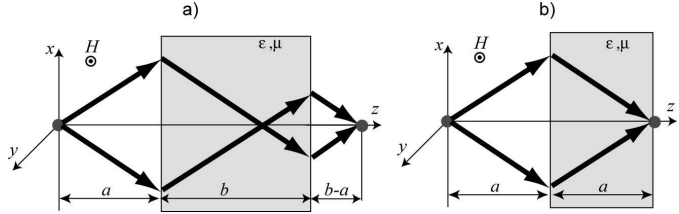,width=3.5in}}
\vspace{10pt}
\caption{Schematic geometry of the generalized planar lens (a), and of the optimal planar lens configuration (b) as described in the text. The image is positioned at $z=2b$; the possible lens configurations span $a\leq b$.}
\label{figConfig}
\end{figure}

Addressing this fundamental question is the primary goal of the present Letter. In contrast to previous analytical work\cite{merlin,podolskiyResolut,shvetsSuperlens}, here we consider a ``generalized'' variant of an imaging system (see Fig.~\ref{figConfig}), where the slab of ``negative'' material with dielectric permittivity $\epsilon=-1+i\epsilon^{\prime\prime}$ and magnetic permeability $\mu=-1+i\mu^{\prime\prime}$ of thickness $b$ is positioned at the distance $a$ away from the object. Note that in contrast to diffraction theory presented in Ref.\onlinecite{efros}, here we are primarily interested in subwavelength resolution. 

We derive an analytical result for resolution of a generalized planar lens, and demonstrate that the resolution is maximized when $a=b$ (the configuration recently implemented in Ref.~\onlinecite{zhang}). We demonstrate that the superlens becomes impractical when $\epsilon^{\prime\prime}\gtrsim 0.3$. Further, we analyze the field distribution in the system and similar to what in the quasistatic limit was proved analytically \cite{MiltonNew} and indicated by other near field investigations\cite{merlin,shvetsSuperlens}, discover a new regime when the EM field has its maxima at {\it both interfaces of NIM}. We show that in contrast to most NIM-imaging descriptions\cite{pendry,smithRESOLUT,webbSuperlens,stockmanSuperlens,podolskiyResolut}, the field structure in generic imaging system is strongly influenced by this new regime, while the conventional ``superlens'' picture (with intensity minimum at the front interface) is rarely realized. Finally, we conclude that the optimal lens configuration suggested in this Letter minimizes the field intensity inside the lens and correspondingly, the total absorption in the imaging system. 

To better illustrate the underlying physics and simplify the analytical results presented in this Letter, we restrict ourselves to the case of imaging of a parallel slot of a thickness $d\ll \lambda$ (with $\lambda$ being the free-space wavelength), extended in the $y$ direction at the origin of a Cartesian system, emitting radiation with TM polarization (Fig.~\ref{figConfig}). The straightforward generalization of results presented here to the case of different shapes of object, lenses, and polarizations will be presented elsewhere\cite{resolutFuture}.

\begin{figure}[t] 
\centerline{\epsfig{file=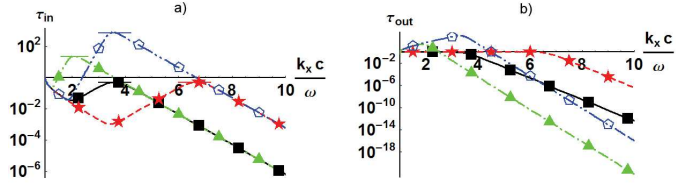,width=3.5in}}
\vspace{10pt}
\caption{(color online) Transfer functions $\tau_{in}$ (a) and $\tau_{out}$ (b) obtained from exact calculations as described in the text (curves), and from Eqs.~(\ref{eqTauIn}-\ref{eqTau}) (symbols); optimal and symmetric lens configurations with various absorptions are shown: black [solid; squares]: $a=b=0.35\lambda$, $\epsilon^{\prime\prime}=10^{-3}, \mu^{\prime\prime}=10^{-6}$; red [dashed; stars] $a=b=0.35\lambda$, $\epsilon^{\prime\prime}= \mu^{\prime\prime}= 10^{-6}$; green [dash-dotted; triangles]: $a=b/2=0.35\lambda$, $\epsilon^{\prime\prime}=10^{-3}, \mu^{\prime\prime}=10^{-6}$; red [dash-dot-dotted; polygons]): $a=b/2=0.35\lambda$, $\epsilon^{\prime\prime}= \mu^{\prime\prime}=10^{-6}$; horizontal lines in (a) correspond to Eq.~(\ref{eqMax})}
\label{figKx}
\end{figure}

The planar-lens imaging can be clearly illustrated in the wavevector space\cite{smithRESOLUT,webbSuperlens,podolskiyResolut}. In this approach, the monochromatic radiation emitted by the source is represented as a series of waves with the same frequency $\omega$ but different wavevectors $k=\{k_x,k_y,k_z\}$. The EM field at an arbitrary point in the system can be calculated as a series of individual waves propagated to this point. Therefore, the problem of imaging an arbitrary source can be reduced to the problem of finding the {\it transfer function} $\tau(x,z;k_x,\omega)$, of an individual wave with fixed ($k_x,\omega$) from the source (origin) to the given point in the system ($x,z$). For TM polarization considered here, it is convenient to work with the $y$ component of the magnetic field: 
\begin{equation}
\label{eqSum}
H_y(x,z;t)=\int a(k_x) \tau(x,z; \omega,k_x) {\rm e}^{-i \omega t} d k_x, 
\end{equation}
where $a(k_x)$ represents the wavevector spectrum of the source [for a very thin source $d\ll\lambda$, $a(k_x)\simeq {\rm const}$]. The transfer function $\tau$ is equal to $\tau_{in}$ inside the lens and $\tau_{out}$ behind it. 

According to the properties of Fourier series\cite{Goodman}, the component with wavevector $k_x$ carries the information about the $x$-structure of a source with a typical scale of $2\pi/k_x$. Therefore, the information about the fine structure of the object is being carried in the waves with $|k_x|\gg \omega/c$. Since $k_x$ and $k_z$ in a plane wave are connected through the dispersion relation $k_x^2+k_z^2=\omega^2/c^2$, these spectral components, also known as evanescent waves, have {\it imaginary} $k_z$ and exponentially decay away from the source. The suppression of the evanescent spectrum is in fact the mechanism behind the resolution limit of an imaging system. 

\begin{figure}[t] 
\centerline{\epsfig{file=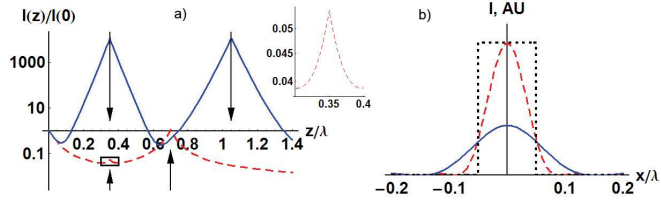,width=3.5in}}
\vspace{10pt}
\caption{(color online) Intensity distributions in symmetric $a=0.35\lambda; b=0.7\lambda$ (blue, solid) and optimal $a=b=0.35\lambda$ (red, dashed) planar lenses with the same absorption $\epsilon^{\prime\prime}=\mu^{\prime\prime}=10^{-6}$. (a) intensity distribution along the focal line $x=0$; note the intensity peaks at both lens interfaces (the relatively small intensity peak at the front interface of optimal system [rectangle] is shown in the inset); upward and downward pointing arrows show the positions of the lens in the symmetric and optimal configuration respectively. (b) Intensity profiles at the focal planes $z=2b$ of imaging systems; dotted black line represents the source; note that the resolution of optimal system is twice better than that of a symmetric structure; resolution of both systems is well-described by Eqs.~(\ref{eqResolut},\ref{eqResolutSer}). }
\label{figFld}
\end{figure}

Thus, the performance limit of a NIM-based planar lens can be related to its transfer function $\tau$ for the evanescent part of the spectrum $(|k_x|\gg \omega/c)$ \cite{podolskiyResolut}. To calculate the transfer function, we first divide the space into three regions: before the lens $(z\leq a)$, inside the lens $(a<z\leq a+b)$, and behind the lens $(z> a+b)$. We then represent the field (of a component with some fixed values of $k_x,\omega$) inside the first region as a sum of incident and reflected waves, the field inside the second region as a sum of transmitted and reflected waves, and the field in the third region as a transmitted wave, and use the boundary conditions to find all coefficients of transmission and reflection. Using this field-matching technique, described in detail in Ref.~\onlinecite{podolskiyResolut}, in the limit of small absorption $\epsilon^{\prime\prime}\ll 1, \mu^{\prime\prime}\ll 1$ for evanescent waves $|k_x|>\omega/c$ we arrive at: 
\begin{eqnarray}
\label{eqTauIn}
\tau_{in}(x,z;k_x,\omega)&=& 
\frac{{\rm e}^{\kappa_z(z-2a)}+i \phi\; {\rm e}^{\kappa_z(2b-z)}}
{(1+i\phi)(1+\phi^2\; {\rm e}^{2 \kappa_z b})}\;
{\rm e}^{i k_x x},
\\
\label{eqTau}
\tau_{out}(x,z;k_x,\omega)&=&
\frac{ {\rm e}^{\kappa_z(2b-z)} }{1+\phi^2\; {\rm e}^{2 \kappa_z b}}\;
{\rm e}^{i k_x x}
,
\end{eqnarray}
where $\kappa_z=\sqrt{k_x^2-\omega^2/c^2}$, and the loss parameter 
\begin{equation}
\label{eqPhi}
\phi=\frac{1}{2}\left[\epsilon^{\prime\prime}+
\frac{\epsilon^{\prime\prime}+\mu^{\prime\prime}}
{2(k_x^2 \;c^2/\omega^2-1)}\right]\ll 1.
\end{equation}
The excellent agreement between the Eqs.~(\ref{eqTauIn},\ref{eqTau}) and the exact solutions of Maxwell equations is shown in Fig.~\ref{figKx}. 

We now derive the resolution limit $\Delta$ of the generalized planar lens. Using the properties of Fourier analysis, the spatial size of a wavepacket at the focal point $(x=0,z=2b)$ (see Fig.~\ref{figConfig}) can be related to its spectral width $\delta$ through the ``uncertainty principle''
\begin{equation}
\label{eqUnc}
\Delta \cdot\delta= 4\pi\xi,
\end{equation}
where constant $\xi\approx 0.6$ depends primarily on the geometry of a source\cite{podolskiyResolut,Goodman}. As seen from Eq.~(\ref{eqTau}), any non-zero absorption in the lens material yields an exponential cut-off of the evanescent spectrum at some critical value of $|k_x^0|=\delta/2$, corresponding to $\tau_{out}(0,2b;k_x^0,\omega)=1/2$. This last relation can be written in the form of a transcendental equation for the resolution limit of a generalized planar lens: 
\begin{equation}
\label{eqResolut}
\frac{2\pi b}{\lambda}=-\frac{\ln\frac{1}{2}\left[
\epsilon^{\prime\prime}+\frac{\epsilon^{\prime\prime}+\mu^{\prime\prime}}
{2 \chi^2}\right]}
{\chi},
\end{equation} 
where $\chi=\sqrt{k_x^{0^2}c^2/\omega^2-1}=\sqrt{\xi^2\lambda^2/\Delta^2-1}$. To determine the optimal configuration of the planar lens, we further simplify the Eq.~(\ref{eqResolut}) assuming that the system has subwavelength resolution $\Delta\ll\lambda$ \cite{footnote1}. In this case $\chi\approx \xi\lambda/\Delta$, and Eq.~(\ref{eqResolut}) yields:
\begin{equation}
\label{eqResolutSer}
\Delta\approx -\frac{2\pi b}{\ln(\epsilon^{\prime\prime}/2)}
\end{equation}
For the case of symmetric planar lens $(b=2a)$ Eqs.~(\ref{eqResolut},\ref{eqResolutSer}) are identical to the ones previously derived in Refs.~\onlinecite{smithRESOLUT,merlin,podolskiyResolut,shvetsSuperlens}. We also note that Eq.~(\ref{eqResolutSer}) describes the resolution of a near-field ``poor-man'' superlens, formed by a planar slab of material with $\epsilon=-1+i\epsilon^{\prime\prime}, \mu=1$\cite{pendry,zhang}. 

One of the main points of this manuscript is to show that the resolution of a planar lens is determined not by the focal distance of the lens $a$, but rather by its thickness $b$, which should be minimized in order to optimize the system resolution (note that the imaging is possible only when $b\geq a$). Thus, for the practical case when the minimum separation between the object and the imaging system $a$ is fixed (for example due to existence of some protective layer, etc.), the best possible configuration corresponds to $a=b$ (see Fig.~\ref{figConfig}b). Note that this particular configuration solves another problem often associated with the superlens -– the local intensity has its maximum exactly at the focal point (as opposed to the symmetric lens configuration, when the intensity maximum at the back interface of NIM region is separated from the focus\cite{pendry}), making it relatively easy to bring the optical system ``in focus''. 

A comparison of Eq.~(\ref{eqResolutSer}) and the resolution of ``conventional'' near-field optics\cite{Goodman} $\Delta_{NF}\simeq 2a$ yields the upper limit for the absorption $\epsilon^{\prime\prime}_{\rm max}$ of practical near-field superlenses when $\Delta=\Delta_{NF}$: 
\begin{equation}
\label{eqAbsLimit}
\epsilon^{\prime\prime}_{\rm max}=2 {\rm e}^{-\pi\xi\frac{b}{a}}
\lesssim 2 {\rm e}^{-\pi\xi}\simeq 0.3. 
\end{equation}
As clearly seen from this last relation, the practical applications of optical sub-diffraction imaging and lithography is limited to non-resonant Ag-based systems, while more absorbing Au-, or Al- based structures, or resonant systems\cite{podolskiyOptExp,soukoulisTHz} will have a resolution below the one achievable via conventional near-field imaging or almost-contact ($a$-separated) lithography. 

Finally, we analyze the intensity pattern and absorption in a planar lens. The field inside the lens region $(a<z\leq a+b)$ is given by Eqs.~(\ref{eqSum},\ref{eqTauIn}). The intensity pattern at the back interface of the lens reflects a well-known effect of evanescent spectrum restoration \cite{pendry,podolskiyResolut}: either constant, or growing exponential (depending on the lens geometry) for the ``weakly-evanescent'' part of a spectrum $\omega/c<|k_x|<k_x^0$, followed by exponential decay for $|k_x|>k_x^0$, where $k_x^0\simeq-\ln(\phi)/b$, as defined by Eq.~(\ref{eqResolut}) (see Fig.~\ref{figKx}). 

In contrast to a widely accepted point of view\cite{pendry,smithRESOLUT}, but in agreement with results in the quasistatic regime\cite{MiltonNew}, the field at the front interface $(z=a)$, may not have a sharp (exponential) minimum even for the case of sub-diffraction imaging. Indeed, as it is clearly seen from Eq.~(\ref{eqTauIn}), there exists some {\it critical wavevector} $k_x^{\rm cr}\simeq -\ln(\phi)/(2b)$, such that the exponential enhancement of the waves with $k_x^{\rm cr}\lesssim |k_x| \lesssim k_x^0$ (Fig.~\ref{figKx}) {\it at the front interface of a lens} is possible (see Fig.~\ref{figFld}). The maximum value of the field can be estimated using 
\begin{equation}
\label{eqMax}
\tau_{in}^{\rm max}=\tau_{in}(0,a;k_x^0,\omega)\simeq (\epsilon^{\prime\prime}/2)^{a/b-1}/2,
\end{equation}
yielding $1/\sqrt{2\epsilon^{\prime\prime}}$ for symmetric lens configuration $(b=2a)$, and $1/2$ for the optimal planar lens $b=a$ described above. Absence of the additional strong field maximum and correspondingly, of the additional absorption associated with such a maximum further illustrates optimality of the design presented here. 

The relation of this phenomenon, which can be attributed to the excitation of coupled surface waves at both sides of a lens (as opposed to an excitation of a surface wave-``anti surface wave'' pair\cite{podolskiyResolut}) to break-up of super-imaging and the onset of the diffraction limit in the system will be described in detail in our future work\cite{resolutFuture}. 

In conclusion, we have developed an analytical approach to the resolution of the generalized planar lens, used this approach to find the optimal (from the resolution standpoint) configuration of the lens system, and derived the maximum acceptable loss in the ``lens'' material in order to achieve a resolution gain over conventional near-field techniques. We also developed an analytical technique to find the field distribution throughout the planar imaging system, and demonstrated that there exists an area of resonant field excitation at the front interface of a lens due to the emergence of a coupled surface wave mode. 

G.W.M. is grateful for support from the NSF through grant DMS-0411035.


\begin{thebibliography}{}
\bibitem{footnote1}Note that for all practical applications $\epsilon^{\prime\prime}, \mu^{\prime\prime}\gtrsim 10^{-6}$ this condition is achievable only in near-field $a,b<\lambda$.
\end{thebibliography}
\end{document}